\documentclass[12pt]{iopart}
\usepackage{textgreek}
\usepackage{textcomp}
\usepackage{graphicx}
\usepackage{url}
\usepackage{iopams}
\usepackage{cite}
\usepackage[colorlinks, citecolor = blue, urlcolor = blue]{hyperref}

\begin{document}

\title{Axion-like particle generation by laser-plasma interaction}

\author{S~Huang$^{1,2}$\footnote{Present address: School of Physics and Astronomy, Tel Aviv University, Tel Aviv-Yafo 6997801, Israel.},
    B~F~Shen$^{1,3,*}$, Z~G~Bu$^{1}$, X~M~Zhang$^{1,3}$, L~L~Ji$^{1}$, S~H~Zhai$^{1,2,3}$}

\address{$^1$ State Key Laboratory of High Field Laser Physics, Shanghai Institute of Optics and Fine Mechanics,
    Chinese Academy of Sciences, Shanghai 201800, China}
\address{$^2$ University of Chinese Academy of Sciences, Beijing 100049, China}
\address{$^3$ Department of Physics, Shanghai Normal University, Shanghai 200234, China}
\begin{indented}
\vspace*{5pt} \item[]$^*$ E-mail: \mailto{bfshen@shnu.edu.cn}
\end{indented}

\begin{abstract}
The hypothetical axion and axion-like particles, feebly coupled with photon, have not yet been found in any experiment.
With the improvement of laser technique, much stronger but shorter quasi-static electric and magnetic fields can be created in laboratory using laser-plasma interaction,
    compared to the fields of large magnets, to help the search of axion.
In this article, we discuss the feasibility of ALPs exploration using planarly or cylindrically symmetric laser-plasma fields as background and an x-ray free-electron laser as probe.
Both the probe and the background fields are polarized such that the existence of ALPs in the corresponding parameter space will cause polarization rotation of the probe, which can be detected with high accuracy.
Besides, a structured field in the plasma creates a tunable transverse profile for the interaction and improves the signal-to-noise ratio via phase-matching mechanism.
The ALP mass discussed in this article ranges from $10^{-3}$ eV to 1 keV.
Some simple schemes and estimations on ALP production and polarization rotation of probe photon are given, which reveals the possibility of future laser-plasma ALP source in laboratory.
\end{abstract}


\submitto{\PS}

\maketitle
%
%

\section{Introduction}
The models of QCD axion and axion-like-particles (ALPs) are a kind of not-yet-found (pseudo)scalar bosons that can potentially and feebly couple with standard model particles like photon \cite{Asano_81, Sikivie_83, Sikivie_85, Bardeen_87, FIP2020}.
The QCD axion \cite{PQ_77, Weinberg_78, Wilczek_78}
    was initially proposed for the strong \textit{CP} problem to balance the charge-parity asymmetry in QCD Lagrangian, and then along with the ALPs became a major candidate for dark matter \cite{Preskill_1983,Dine_1983,Abbott_1983, Covi_99, Duffy_2009, Sikivie_10, Masso_08,BraatenZhang_axionstar}.
In the minimal axion models like the KSVZ model \cite{Kim_79, SVZ_80},
    one axion couples with two photons.
Viewing from a macroscopic perspective, one can treat this axion-diphoton (``$a\gamma\gamma$'') process as a three-wave mixing.

Several detection schemes based on the ``$a\gamma\gamma$'' process have been proposed or carried out. For example,
helioscopes with meter-long magnets have been built and planned to detect possible axion flux from the sun \cite{CAST_05, CAST_07, CAST_09, CAST_17, IAXO_14, IAXO_19}.
Optical or microwave cavities have been used to amplify the weak signal of converted photons \cite{Bradley_rev03, Melissinos_09, ADMX_10, ADMX_20, Lasenby_20, Sikivie_14}.
ALP creations in laboratories, which are independent of possible natural ALP sources, have been proposed and conducted including the light-shining-through-wall (LSW) experiments \cite{vanBibber_87, Ruoso_92,Cameron_93, GammaV_08, Afanasev_08, ESRF_08,ESRF_10, OSQAR_08,OSQAR_13, ALPS_09,ALPS_10,ALPS_13,ALPS_15, LUXE_a21}
    where probe photons are expected to regenerate after converting into ALPs and traveling through a photon-blocking wall.
The change of the state of probe photon after interaction is also a proof of ALP's existence \cite{PVLAS_06,PVLAS_e07,Ahlers_07,PVLAS_16,PVLAS_2020, Tam_12}.
For XFEL beam with very narrow frequency bandwidth, multiple reflections inside a crystal can be used to detect the beam's ellipticity with a precision of $|\epsilon_\mathrm{min}|^2 \approx 10^{-12}$ \cite{marx-crystal11,marx-crystal13, Schlenvoigt2016,Schulze2018,Bernhardt2020,Schmitt_21,Schulze2022}.

In the context of the ``$a\gamma\gamma$'' process, the ALP models are distinguishable from each other by a few key parameters:
    the particle's mass $m_a$ and decay constant $f_a$,
    and the coupling coefficient $g_{a\gamma\gamma}$.
The astrophysical and cosmologic observations and the null-result experiments constrained the QCD axion's mass in $10^{-6}$ to $10^{-2}$ eV.
The recent CAST result limited $g_{a\gamma\gamma} < 6.6 \times 10^{-11}\ \mathrm{GeV}^{-1}$ for $m_a < 0.02\ \mathrm{eV}$ \cite{CAST_17}.
Those results are usually shown in a $g_{a\gamma\gamma}$ versus $m_a$ plot (like \fref{fig:alplot}).
Beyond the minimal invisible ALP models, series of pseudoscalar or scalar ALP models with wide mass ranges, including the ultralight ALP as dark energy \cite{Choi2021}, were proposed.
In this article, pseudoscalar ALPs with mass from $10^{-3}$ to $10^{3}$ eV are investigated.

To enhance such a weak interaction, one has to either elongate the effective interaction length or intensify the interaction.
Recently, the electromagnetic fields of ultrafast-superstrong lasers, along with the quasi-static electric or magnetic fields created in laser-plasma interactions have attracted researcher's attention \cite{Gies_2009jhep, Dobrich_2010, VillalbaChavez_14, VillalbaChavez_2017, Burton_18, Tercas_2018, Lawson_19, Mendonca_20}.
Laboratory laser-plasma acceleration of charged particles \cite{Tajima_79, Clark_00, Pukhov_01, Shen_01} has made its success.
The accelerating electric field can reach as high as $10^{12}\ \mathrm{V}\ \mathrm{cm}^{-1}$ using the available or under-construction 10 or 100-PW laser \cite{vulcan, apollon, eli, sacla, shine},
    much stronger than a state-of-the-art hundred-tesla magnet (equivalent to $10^9\ \mathrm{V}\ \mathrm{cm}^{-1})$.
With such a field strength, extremely high-intensity lasers have their unique potential in the ``table-top'' ALP search.
Moreover, the dispersion in ``$a\gamma\gamma$'' three-wave mixing can improve the signal-to-noise ratio \cite{Lawson_19, Mendonca_20}.
The existence of ALP's mass has made a nuisance in the detectors based on ``$a\gamma\gamma$'' process because photons are massless.
In plasma or other material that structures the laser's field or introduces a dispersion term, the phase of ALP during the interaction can be tuned and matched and the signal therefore enhanced.
In addition, the created ALP pulse should have the same timespan as the ultrafast laser as the stimulating background field.
Via coincidence detection, one can significantly suppress the white noise by only taking data in a short temporal window.

It is worth mentioning that the ``$a\gamma\gamma$'' process and vacuum light-by-light scattering and birefringence \cite{EH36, Schwinger51, Klein64, BB70}
    are similar in many ways.
One of the reasons is that their Lagrangians have alike factors, although the simplest vacuum light-light scattering is a process of nonlinear four-wave mixing.
Some detailed discussions gave their estimates of the ALP influence on the vacuum QED experiments \cite{Wilczek_87, Gies_2009epjd, Tommasini_2009, VillalbaChavez_13, VillalbaChavez_18,sacla,PVLAS_2020,Fedotov2022}.
A kind of ALP-regeneration experiments called ``axion four-wave mixing'' were also conducted \cite{Hasebe_15, Homma_17, Nobuhiro_20}
    where ALP was supposed to work in the middle as the electron-positron pair in QED vacuum four-wave mixing.

In this article, we investigate the generation of ALP's flux in a three-wave mixing between the strong quasi-static electric background fields in plasma and a x-ray free-electron laser (XFEL) probe field \cite{EUXFEL2017,Lu2018,SHINE_XFEL2014,Yan2019,Huang2020}.
The calculation is based on the ``$a\gamma\gamma$'' process that treats ALP field as a scattering material for light field and utilizes electromagnetic field as the source term in ALP's wave equation.
Two geometric designs of the background static electric fields are examined.
The first is a planar model where the electric field is built between two electrode planes, which appears in the laser-plasma target normal sheath \cite{Clark_00, Pukhov_01}
    or a light pressure compressed foil \cite{Shen_01}.
The second is a model with cylindrical symmetry which can be found inside a laser-plasma wakefield \cite{Tajima_79}
    or around a well-collimated charged particle beam.
We will give details of ALP-photon interaction models in \sref{sec:model}, and
    proceed with applications of the models in field formation, detailed estimations, and discussions and in \sref{sec:scheme}.
Although we base our calculations on specific kinds of field structures, the methods can be applied to other kinds of designs, such as changing the probe's waveband from x-ray to microwave or visible light.

\section{Models} \label{sec:model}

Pseudoscalar ALP models has a Lagrangian density of the ``$a\gamma\gamma$'' process as
\begin{equation} \label{eq:agglagrangian}
    \mathcal{L} = -\frac{1}{4}g\tilde{F}_{\mu\nu}F^{\mu\nu}\psi.
\end{equation}
For the sake of simplicity, the subscript under the coupling coefficient $g_{a\gamma\gamma}$ in following sections is omitted, and
the natural Lorentz--Heaviside units are used as $c=\hbar=\epsilon_0=1$.
Symbols $F$ and $\tilde{F}$ stand for the normal and dual electromagnetic field tensors;
$\psi$ is the ALP's field.
\Eref{eq:agglagrangian} predicts the generation of ALP as a result of Klein--Gordon equation for spin-zero particles with a stimulating source proportional to the anisotropy invariant of electromagnetic field
\begin{equation}
    (\partial^2_t-\mathbf{\nabla}^2+m_a^2)\psi = g \bi{E}\cdot\bi{B}. \label{eq:kleingordon}
\end{equation}
The detailed solving steps to \eref{eq:kleingordon} can be found in \ref{sec:appendix1}.

\subsection{Planar background field}\label{sec:planar}

\begin{figure} [hbtp]
\centering
\includegraphics[width=0.9\textwidth]{Figures/figure_1abc}
    \caption{\label{fig:1d}
    Sketches of the interaction between a magnetically $y$-polarized probe and electrically (a) $y$-polarized or (b) $z$-polarized background field.
    And (c) the Feynman diagram for ALP creation for this scheme.
    The vertical separation of ALP and probe beam in (a)'s right part is only for the benefit of observation and does not suggest a spacial separation along $z$ direction.
}\end{figure}

Shown in \fref{fig:1d} as a simplified view, in a one-dimensional interaction, a probe (``pb'') photon annihilates while travelling in a parallel-polarized background (``bg'') field and a spinless ALP is created.
Under the slowly-varying envelope approximation (SVEA), we treat the probe laser beam as a plane wave transmitting along the longitudinal $x$-axis with a magnetic field like
\begin{equation}
    \bi{B}_{\mathrm{pb,I}} = B_0 \hat{\bi{e}}_\mathrm{pb}
        \exp{\left\{-i(\omega t - k x)\right\}}. \label{eq:pb1}
\end{equation}
The vector $\hat{\bi{e}}_\mathrm{pb}$ of probe in \eref{eq:pb1} indicates the probe's magnetic polarization which can be $y/z$-linear (like in \fref{fig:1d}), or other combinations of independent polarization pairs such as left/right-circular and azimuthal/radial polarised.
Both the plane-wave assumption and SVEA are pretty strong constraints.
For instance, a tightly focused laser beam can only be treated as a plane wave within one Rayleigh range from the focal point.
The ALP, however, is not necessary a planar wave \cite{Gies_2009jhep, VillalbaChavez_2017} if the minimal transverse length scale is too short.
But the de Broglie wavelength of ALP is long enough so the effectiveness of conversion is not influenced due to diffraction.
Since the post-generation ALPs have very limited influence on the probe, it is reasonable that we use the planar assumption to calculate the loss of photons.

The other source factor in \eref{eq:kleingordon} is the background electric field. It can either (I) be uniform (in the interaction area) or (II) have designed transverse structure.
For the first model, the electric field goes as $\bi{E}_\mathrm{bg,I} = E_0 \hat{\bi{e}}_\mathrm{bg}$. We denote
\begin{equation} \label{eq:pfactor}
    p = \hat{\bi{e}}_\mathrm{pb} \cdot \hat{\bi{e}}_\mathrm{bg}
\end{equation}
as polarization factor, which will be a constant if the probe and background fields are of the same polarization pair mentioned above.
Under these assumptions, the three-wave-mixing of ``$a\gamma\gamma$'' process gives an axion's wave function as
\begin{equation}
    \psi_\mathrm{I} = \frac{i g p E_0 B_0 L}{2 k_a}
        \exp{\{-i(\omega t - k_ax)\}}\,\mathrm{sinc}\left\{
            \frac{1}{2}(k-k_a)L
        \right\}. \label{eq:axion1}
\end{equation}
The result in \eref{eq:axion1} is an accumulation of ALP during a planar probe beam transmitting for a length $L$ along the $x$-axis in a uniform background.
The amplitude of the stimulated ALP field is inversely proportional to the ALP's momentum $k_a = \sqrt{\omega^2-m_a^2}$ when the ALP's energy is assumed equal to the probe photon's.
Ignoring the phase mismatching, one can obtain the ALP production ratio in this process as
\begin{equation}
    \frac{N_\mathrm{I}}{N_\mathrm{pb}} = \frac{1}{N_\mathrm{pb} m_a |\psi_\mathrm{I}|^2 V}
        = \frac{1}{2} m_a g^2 p^2 \frac{E_0^2 L^2 \omega}{k_a^2}  \label{eq:naxion1}
\end{equation} which is equal to the probe photon's loss.
In \eref{eq:naxion1}, $V$ stands for the ``volume'' of stimulated ALP beam that has an approximately same size of the probe's, and hence $N_\mathrm{pb}=(B_0^2 V)/(2\omega)$ is the probe photon number.
The ALP production ratio in a one-shot interaction is determined by several terms:
(i) the phase-matching effect,
(ii) the ALP parameters,
(iii) the background electric field and length, and
(iv) the probe's frequency.

\begin{figure}[htbp]
\centering
\includegraphics[width=0.9\textwidth]{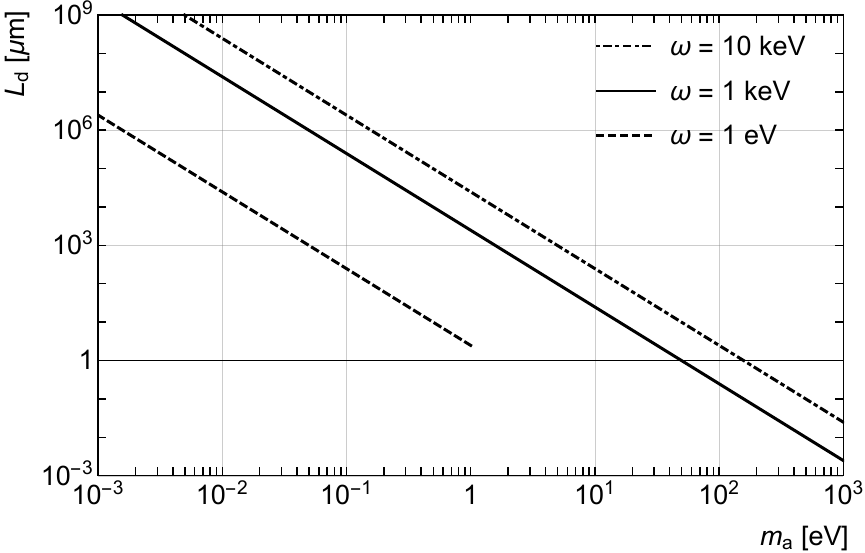}
    \caption{\label{fig:dephasing}
    Dephasing lengths for planar 1-eV optical laser (dashed line), 1-keV (solid line), and 10-keV (dot-dashed line) XFELs as probes in the ``$a\gamma\gamma$'' three-wave-mixing process for different ALP masses. The phase-matching effect can be neglected in short interactions with light ALPs where the interaction length $L<L_\mathrm{d}$.
}\end{figure}

The sinc factor in \eref{eq:axion1} describes the phase-matching effect. It gives a dephasing distance
\begin{equation} \label{eq:dephasing}
    L_\mathrm{d} = \frac{2\pi}{k-k_a} \approx \frac{4\pi\omega}{m_a^2}
\end{equation} if $\omega \gg m_a$.
Taking $\omega = 1\ \mathrm{eV}$ and $m_a = 10^{-3}\ \mathrm{eV}$, one can find the distance to be 2.5 meters, which is omittable in the micro-to-centimetre-long strong-field laser-plasma experiments.
But for heavier ALPs above $10^{-1}\ \mathrm{eV}$, the distance is shortened to under several hundred micrometers.
In the case when $L\gg L_\mathrm{d}$, the production rate will be capped at
\begin{equation} \label{eq:fulldephase}
    \frac{N_\mathrm{dephased}}{N_\mathrm{pb}}
        = \frac{1}{2} m_a g^2 p^2 \frac{E_0^2 L_\mathrm{d}^2 \omega}{k_a^2}.
\end{equation}

In an experiment looking for the change of probe's polarization state, the axion production ratio $N_\mathrm{I}/N_\mathrm{pb}$ is proportional to the square of ellipticity or polarization rotation~$|\epsilon|^2$.
For instance, if the direction of background field is set as $\hat{\bi{e}}_\mathrm{bg} = \hat{\bi{e}}_y$ and the probe is 45\textdegree-linear polarized as $\hat{\bi{e}}_\mathrm{pb} = (\hat{\bi{e}}_y \pm \hat{\bi{e}}_z)/\sqrt{2}$,
then only the probe's component with $y$ polarization contributes to the ALP production and losses photons, leaving the $z$-polarized unaffected.
In this case, after the ``$a\gamma\gamma$'' process a polarization rotation angle of
\begin{equation}
    \epsilon = \left(\frac{N_\mathrm{I}}{N_\mathrm{pb}}\right)^{\frac{1}{2}}
        \approx \left(\frac{m_a g^2}{2\omega}\right)^{\frac{1}{2}} p E_0 L \label{eq:ellip1}
\end{equation}
will be generated.
The alike relations also apply for the other pairs of polarization combinations.

\subsection{Cylindrical background field} \label{sec:2b}
The model with one-dimensional laser-plasma field sets a benchmark of ALP production as in a near-field three-wave-mixing process in \eref{eq:naxion1} and \eref{eq:ellip1}.
The model in this subsection, however, focuses on the detection of a particular kind of ALP with certain mass by extending the search range or improving the signal-to-noise ratio in a particular band.
In this model, a cylindrically symmetric electric field $\bi{E}_\mathrm{bg,II} = E_0 f_E (r){\hat{\bi{e}}}_\mathrm{bg}$ interacts with
    the probe's magnetic field $\bi{B}_\mathrm{pb,II} = B_0 f_B (r) {\hat{\bi{e}}}_\mathrm{pb} \exp{\left\{-i\left(\omega t - k_x x\right)\right\}}$.
The radial distribution functions of the electric and magnetic fields are $f_E$ and $f_B$, respectively. The interaction is corresponding to their product $f(r)=f_E f_B$.
Due to the possible existence of radial momentum, the probe photon's longitudinal momentum $k_x$ is unnecessarily equal to its energy $\omega$.
Under similar assumptions and methods, the ALP's wave function is obtained as
    $\psi_\mathrm{II}(x,r,t) = \sum_{m}\psi_{\mathrm{II}m}$ where
\numparts
    \begin{eqnarray} \label{eq:axion2}
    \fl
    \psi_{\mathrm{II}m} =
        \frac{igpE_{0} B_{0} L}{2k_a} \exp{\{-i(\omega t - k_a x)\}} \,
        \mathrm{sinc}\left\{
            \frac{1}{2} \left(k_{x} - k_a + \frac{k_{m}^{2}}{2k_a} \right) L
        \right\} f_{m} \mathrm{J}_{0}(k_{m} r),
    \end{eqnarray} \begin{eqnarray} \label{eq:fm}
    f_{m} = \frac{2}{\rho^{2}[\mathrm{J}_{1}(k_{m}\rho)]^{2}}
        \int_{0}^{\rho}{f(r) \mathrm{J}_{0}(k_{m}r) r\mathrm{d}r}.
    \end{eqnarray}
\endnumparts
The ALP's transverse component $\bi{k}_a$-related momentum is defined as $k_m = u_m / \rho$, where $u_m$ is the $m$th zero point of Bessel $\mathrm{J}_0$ function, and $\rho$ is the radial boundary of interaction region.
Calculation details are in \ref{sec:appendix1}.

The second model leads to a change in the phase-matching term.
Specifically, in a short interaction with length $L \ll L_\mathrm{d}$, the two models give the same result.
But with longer $L$ near or over the plane-wave dephasing distance, only the $\psi_{\mathrm{II} n}$ components who satisfies
\begin{equation}
    k_{x}-k_a-\frac{k_{n}^{2}}{2k_a} = 0 \label{eq:pmcondition}
\end{equation}
can survive and the others vanished.
Besides in \eref{eq:axion1}, the phase-matching effect always reduces the ALP's intensity because $k=\omega>k_a$.
Predicted by \eref{eq:pmcondition}, a layout with radial structure has a longer dephasing length around a selected ALP mass that is determined by the radial profile function $f(r)$.
The selective effect benefits the production of heavier ALP with mass close to the energy of probe $\omega$.
The ALP production ratio that corresponds to the $f_n$ component is
\begin{equation}
    \epsilon^2_n = \frac{N_{\mathrm{II}n}}{N_\mathrm{pb}} = \frac{2V}{\rho^{2} N_\mathrm{pb}}
        \int_{0}^{\rho}{r\mathrm{d}r m_{a}|\psi_{\mathrm{II}n}|^{2}}
    =   \frac{N_\mathrm{I}}{N_\mathrm{pb}} |f_{n}|^{2}[\mathrm{J}_{1}(k_{n}\rho)]^{2},
\end{equation}
a result proportional to \eref{eq:naxion1}.
Eventually, when $L \gg L_\mathrm{d}$, all components are dephased (due to the limit $\omega \ge k_n$) to \eref{eq:fulldephase} indicating that further elongation of interaction length will no longer benefit ALP production rate.


\section{Detection schemes} \label{sec:scheme}
In this section, we are discussing two experimental schemes based on the two models in \sref{sec:model}, using the key parameters of the upcoming Shanghai High-Repetition-Rate XFEL and Extreme Light Facility (SHINE).
SHINE is going to have a MHz XFEL generating $10^{9}$ to $10^{14}$ of 0.4 to 25 keV photons every bunch \cite{SHINE_XFEL2014}, while its 100-PW laser will have 15 fs pulses at 910 nm wavelength \cite{shine}.
The optical laser is used to create a quasi-static background electric field in plasma, and the XFEL works as a probe for the ALP search.
Disturbance in the laser-plasma interaction that harms the experimental uncertainty is temporarily laid aside and will be discussed in \sref{sec:discussions} and \ref{sec:appendix2}.

\begin{figure}[htbp]
\centering
\includegraphics[width=.45\linewidth]{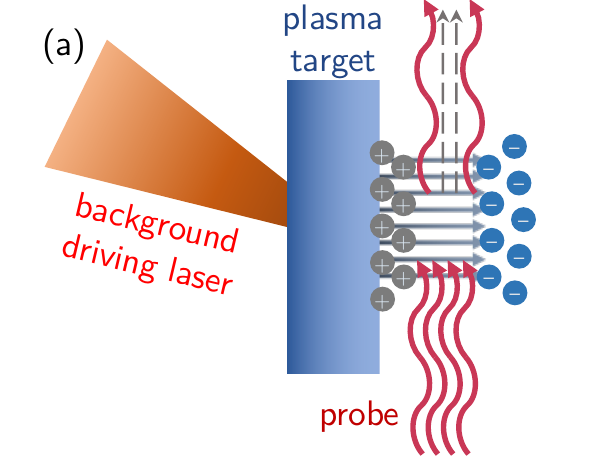}
\includegraphics[width=.45\linewidth]{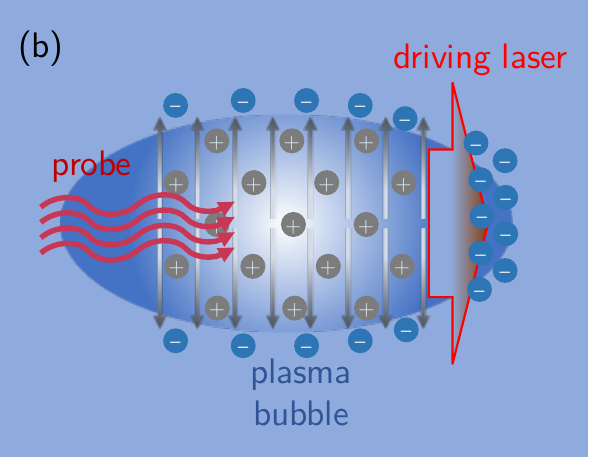}
    \caption{\label{fig:scheme}
Sketches of the schemes of (a) planar electrodes and (b) cylindrical bubble.
In (a), the driving laser shots on an overdense plasma target and then an electric field is formed behind the target, while a probe beam goes through the field and some photons turn into ALPs. In (b), the driving laser shots into an underdense plasma and a bubble of electron density is created. The probe which follows the driving laser will be influenced by the charge-separation field in the bubble where some probe photons turn into ALPs.
}\end{figure}

\subsection{Electric field in overdense plasma electrodes} \label{sec:scheme_1}
As shown in \fref{fig:scheme}(a), when a high-power laser hits on an overdense plasma target, electrons are pushed out by the pondermotive force, and
    the ions, relatively heavier, do not react instantly but remain in their original positions.
For the time being, between the pushed electrons and the separated ions in the target, an electric field builds up.
If the laser is linearly polarized and the target is thick, only a part of the electrons can be pushed out and form a gradient field that is also known as the target-normal-sheath field (TNSF).
And when the laser is circularly polarized and hits the target at a right angle, almost all electrons will be pushed out.
These fields have been used to accelerate ions from tens-of-MeV to sub-GeV.
Between the two group of charged particles, the field between electrodes has an estimated strength
\begin{equation}
    E_0 = Q/S.
\end{equation}
If a charge of 10 nC in the cross section $S=10\ \mathrm{\mu m}^2$ is pushed out, the field can reach $E_0 = {10}^{12}\,\mathrm{V}\,\mathrm{cm}^{-1}$ in order of magnitude.
The interaction length should be $L=\sqrt{S}=3.2\ \mathrm{\mu m}$, correspondingly.
With this setup and $10^{14}$ 1-keV photons from SHINE, if we require $\epsilon^2 > \epsilon_\mathrm{min}^2=10^{-12}$, according to \eref{eq:naxion1} and \eref{eq:ellip1}, the qualified ALP must have
\begin{equation}
    m_a g^2 > 1.8\times {10}^{-9}\ \mathrm{eV} \ \mathrm{GeV}^{-2}.
\end{equation}
Self evidently, the ALP mass should not exceed the probe's energy 1 keV.
The dephasing takes place at $m_a > 28\ \mathrm{eV}$, resulting in a gradual drop of ALP production.

\subsection{Electric field in underdense plasma bubble}
The second scheme [\fref{fig:scheme}(b)] is based on a laser-driven hollow structure in plasma.
When a short-pulse laser beam enters an omnipresent low-density target, its pondermotive force pushes electrons ahead and excites a plasma wave.
The pushed electrons are flowing backwards forming a wakefield behind the laser.
In this ploughing process, ions are almost immobile due to their heavy masses, which causes an electron density ``bubble'' blowing in the wakefield between the accumulating electrons at the head of the laser and flowing back electrons at the tail.
The size of the bubble $R$, if driven by optical laser, is usually around a few to tens of micrometres.
And the field along the axis is 0 due to the symmetry.
To make use of the bubble field, the XFEL probe with size $\rho$ of few to tens of nanometres aims at the edge where the radius $r_0 < R$.
Here, the polarization of the XFEL should be adapted to the radial polarization of background field in the bubble.

The local static electric field has an amplitude
\begin{equation}
    E_{0}=\frac{e}{2}n_\mathrm{pe}r_0. \label{eq:bubbleedge_profile}
\end{equation}
Symbol $e$ stands for the electric charge of an electron.
The plasma density $n_{\mathrm{pe}}$ should be near or lower than the critical density $n_{\mathrm{cr}}=\omega^2 m_e/(4\pi\alpha)$ which is $1.4\times{10}^{21}\ {\mathrm{cm}}^{-3}$ for 1.4-eV driving photons (whose wavelength is 910 nm) and corresponds to the maximum of $E_0 = 1.3\times{10}^{12}\ \mathrm{V}\ \mathrm{cm}^{-1}$ (in a case where the bubble's radius $R=10\ \mathrm{\mu m}$).
It is possible that a laser wakefield bubble transmits within plasma for tens of centimetres~\cite{Palastro_2020},
    although creating a bubble in the near-critical-density plasma requires a very strong driving laser to build relativistic transparency for electrons.
We pick an interaction length $L = 1\ \mathrm{cm}$.
For the same probe setting as in \sref{sec:scheme_1} assuming $f_B=1$, near the bubble edge ($r_0\approx R$) and before any dephasing, the ALP parameters are bound to
\begin{equation}
    m_a g^2 > 5.4\times {10}^{-17}\ \mathrm{eV} \ \mathrm{GeV}^{-2}
\end{equation}
to achieve $\epsilon^2 > \epsilon_\mathrm{min}^2$ constraint.
Using \eref{eq:dephasing}, one can estimate the dephasing to happen when $m_a$ reaches 0.5 eV for 1-keV probe and 1-cm interaction length in the one-dimension model, very close to $k_1$ in \tref{tab:km} of the cylindrically symmetric model.

\begin{figure}[htbp]
\centering
\includegraphics[width=.9\linewidth]{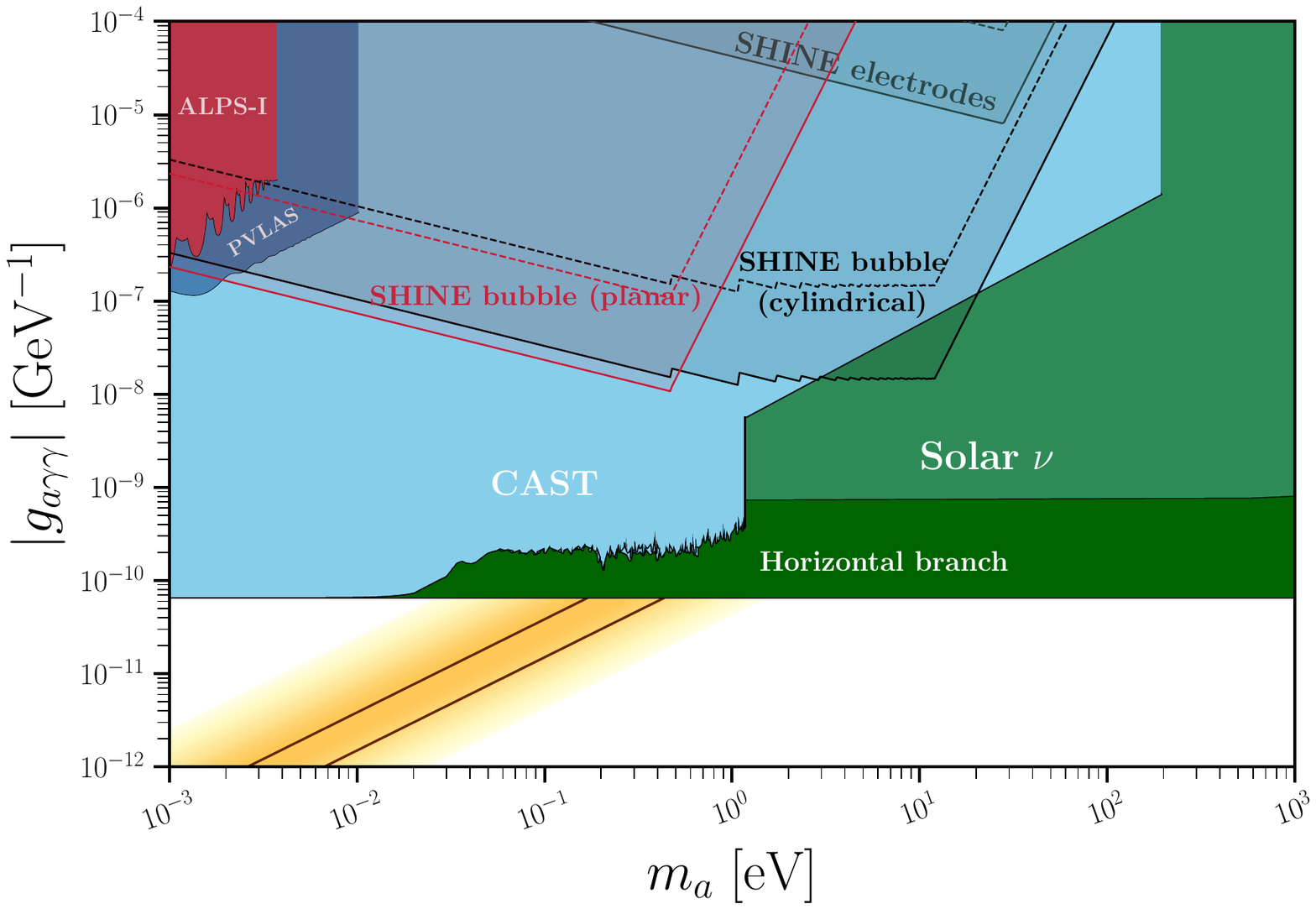}
    \caption{ \label{fig:alplot}
    An ALP parameter space for various experiments (labelled in white letters) and the two schemes estimated using SHINE's setup with 1-keV XFEL (label-led in dark letters) when $\epsilon^2>\epsilon^2_\mathrm{min} = 10^{-12}$. The dashed lines are the ones with $\epsilon^2>10\epsilon^2_\mathrm{min} = 10^{-11}$ condition. The yellow region stands for the QCD axion models. \newline This figure is generated with codes in \cite{axionlimits}.
}\end{figure}

The other idea is to enlarge the XFEL beam to fit the bubble size, namely let $\rho = R$. In a simplified two-dimensional cylindrical model of wakefield, the static electric field in the bubble has an amplitude and radial profile as
\begin{equation}
    E_{0}=\frac{e}{2}n_\mathrm{pe}R,\quad f_{E}(r)=\frac{r}{R}. \label{eq:2dprofile}
\end{equation}
To use the radial background field, a azimuthally polarized probe beam is needed \cite{Morgan2020}.
For the same bubble, the response in \eref{eq:fm} shows a series of peaks on the $k_n$-spectrum.
The first five coefficients are listed in \tref{tab:km}.
Before any dephasing, the ALP parameters under $\epsilon^2 > \epsilon_\mathrm{min}^2$ should satisfy
\begin{equation}
    m_a g^2 > 1.1\times {10}^{-16}\ \mathrm{eV} \ \mathrm{GeV}^{-2},\quad
    \omega < 0.47\ \mathrm{eV}
\end{equation}
However, the phase-matching effect given by \eref{eq:pmcondition} varies for different $k_n$.
At 0.47 eV, the first-order component dephases and is taken out from the equation, and so forth for the rest of high-order components at different points.
Then all components dephase when $k_n \ge \omega$.
From \fref{fig:alplot}, it is clear that the bubble scheme goes through a longer and slower dephasing process than the planar one due to the two-dimensional phase-matching effect.
Beside the cylindrically symmetric scheme, other fields with various complex transverse structures or vectorial polarization can provide similar effect.

\begin{table}
\caption{First five transverse momentum $k_n$ and the corresponding production ratio for the setup with 1 keV probe, $\rho=10\ \mathrm{\mu m}$, and $L_\mathrm{d}=1$ cm.}\label{tab:km}
\begin{indented}
\item[]\begin{tabular}{@{}lll}
\br
$n$ & $k_n$ [eV]    & $N_{\mathrm{II}n}/N_{\mathrm{I}}$\\
\mr
1   & 0.47         & 0.18 \\
2   & 1.1         & 0.15 \\
3   & 1.7         & 0.047 \\
4   & 2.3         & 0.030 \\
5   & 2.9         & 0.017 \\
\br
\end{tabular}
\end{indented}
\end{table}

\subsection{Further discussions} \label{sec:discussions}
Although this paper is not designated to the technical issues in a possible future experiment, some preliminary discussions will follow in this subsection.
The experimental design will be addressed in details by following research works.

The schemes described in this paper are to detect the photon number of a collimated and polarization-purified probe beam in x-ray waveband, typically, around 1 to 10 keV.
The noise that can deteriorate resolution of the detector, namely a x-ray polarization analyser, should be the photons with similar momentum.
The resolution is benefited from the very design of the polarizer and analyser \cite{Bernhardt2020,Schulze2022,Schulze2018} which work through multiple reflections on crystal surfaces.
Due to dispersion, the crystal only functions properly in a very narrow bandwidth from $10^{-2}$ to 1 eV depending on the desired polarisation resolution/purification.
And their emittance/acceptance angles are also limited around milliradians, which is a match to the character of the XFEL \cite{SHINE_XFEL2014,Yan2019,Huang2020}.
The polarized XFEL probe with over $10^{10}$ photons also has a good capability to resist unpolarized noise.
The false signal, given to be unpolarized x-ray photons, will be added statistically uniformly to both polarization directions.
Only when the pollution from unpolarized noise photons has the same amount of the probe's photons, the polarization angle deviated from the neutral state, 45{\textdegree} in linear polarized setup for instance, will be halved.
The possible noise sources are discussed in \ref{sec:appendix2}.

\begin{figure}[htbp]
\centering
\includegraphics[width=.5\linewidth]{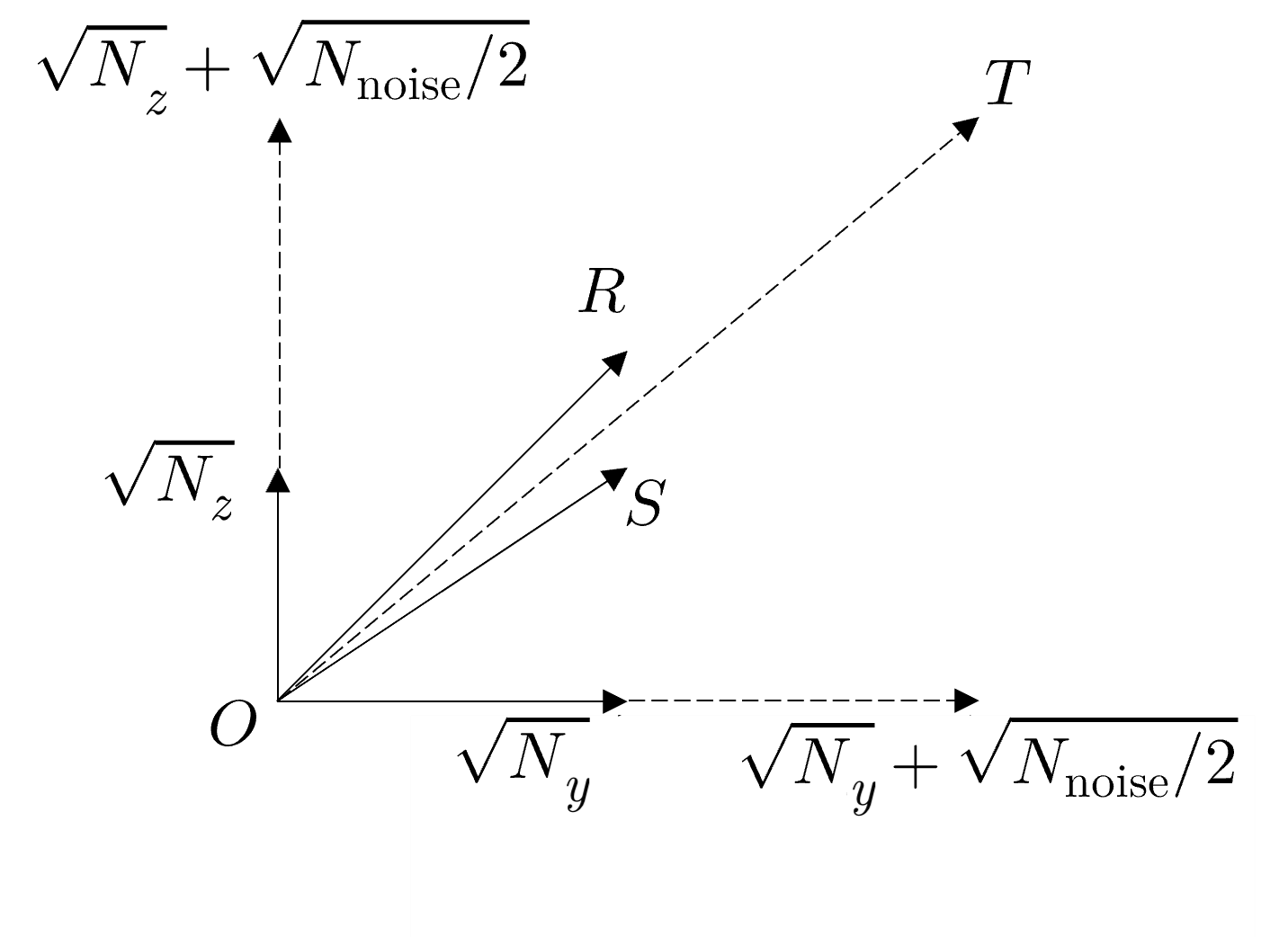}
    \caption{ \label{fig:signal_noise}
    The polarizations of the original probe (vector $OR$ with angles $ROy = ROz = 45$\textdegree), the probe losing $z$-polarized component after the interaction (vector $OS$), and the after-interaction probe with unpolarized noise (vector $OT$). The horizontal and vertical axes indicate the amplitudes of light's two polarization components. The original probe has $2 N_y$ photons.
    Only if $2 N_y = N_\mathrm{noise}$, the deviation angle with noise $TOR$ will be the half of the signal's $SOR$.
}\end{figure}

Spatial-temporal synchronizations are also crucial for a successful experiment.
In the electrode scheme, the charge-separation field can last up to nanoseconds before the ions are dragged out and fully neutralize the field.
While in the bubble scheme, the probe has to be injected in tens of femtoseconds after the driving laser.
The recent progresses in \textmu m-fs synchronization can be found in \cite{Nakatsutsumi2016,Gaus2021,Xin2018}.
The plasma target preparation also constrains experimental repetition rates.
Solid-state targets can be provided with higher rates than the gas targets.
Additionally, the interaction chamber for gas plasma needs to be regularly revacuumized and limits the repetition rate.
For recent progresses in improving the repetition rate in laser-plasma interaction experiments, please refer to \cite{Prencipe2017,Qin2022,Poole2018} for solid target, \cite{Morrison2018} for liquid target, and \cite{Dann2019} for gas target with a active feedback system to control the stability of laser.
For SHINE, the future ALP search experiment would be able to run at a rate of $10^{-3}$ to 1 Hz.
It is also possible to produce a sequential wakefield bubbles using one pumping laser pulse \cite{He2013}, and inject multiple probes with different delays \cite{Lu2018,Yan2019} into each bubble.
To exploit the whole bubble, several XFEL beams with different delays and polarizations can also be guided into the same bubble at various positions in polar direction.
Although this will require a spatial synchronization at nanometre scale.

Worth mentioning is the progress of another similar scheme with a collimated charged particle beam (e.g. the GeV-nC electron beam in the European XFEL \cite{EUXFEL2017}).
The electric field around the particle beam has a cylindrically estimated form like $E_{0}= Q/(2\pi\rho\gamma L) \quad f_{E}(r)=\rho/r$.
In a head-to-head collision between 10 nC electron beam and a polarized 1-keV probe photon beam, a rough calculation using the Lorentz-invariance of \eref{eq:kleingordon} gives a bound $m_a g^2 > {10}^{-13}\ \mathrm{eV} \ \mathrm{GeV}^{-2}$ on $m=1$ band for $\epsilon^2 > \epsilon_\mathrm{min}^2$.
The XFEL facility can both provide the electron beam and x-ray photon beam.
The lower result is because of the much weaker field around the electrons than the one in laser-plasma bubble.
But the gains are lower disturbance, easier to synchronize, and higher repetition rate.

Lastly, although this article is based on keV XFEL photon, optical or microwave photons may also be used as polarization probes.
Unlike the vacuum birefringence experiments, the ellipticity change or polarization rotation caused by ALPs are inversely proportional to the probe's frequency.
They have much higher $N_\mathrm{pb}$ and easier to be tuned than the XFEL, though
the polarization purification and detection techniques in visible light (PVLAS at about $10^{-8}$ at few Hz \cite{PVLAS_2020}) and microwave \cite{Zhang_15, Kimball_17} are not as good as in the x-ray, and the effective interaction length will be limited with a much shorter dephasing length, as shown in \fref{fig:dephasing}.
On the other end, probe with higher frequency can help the search of ALP with higher mass. The reason we choose 1-keV XFEL as the probe is because it has enough amount of photons per shot that $N_\mathrm{pb}\ge 1/{|\epsilon_\mathrm{min}|^2}=10^{12}$.
This limit would also be satisfied for 10-keV XFEL at SHINE, in which case the ALP yields in \eref{eq:naxion1} will drop but the dephasing distance in \eref{eq:dephasing} will elongate, resulting a right shift of the SHINE bounds in \fref{fig:alplot} from $10^0$-eV to $10^1$-eV region.
In another words, the strength of the background field will determine how weak can the search go along the coupling-coeffecient dimension, while the ALP's mass dimension depends on the probe's frequency and the transverse profiles.

\section{Conclusion}
In this article, we explored the feasibility of using the electromagnetic fields created in laser-plasma interactions to detect axion-like particles in laboratory.
The $10^{12}\ \mathrm{V}\ \mathrm{cm}^{-1}$ background fields are going to be generated by one of the strongest laser around the world in the under-construction SHINE facility.
A beam of coherent x-ray by XFEL is employed to probe the background, and the probe and the background are polarized with an angle of 45{\textdegree} in between.
If ALPs do exist in the corresponding parameter space, a change of the probe's polarization state will be found.
The calculations through the Lorentz-invariant axion-diphoton process were shown in two different models.
With the key parameters of SHINE, the related ALP parameter spaces were given as $m_a g^2$ varies from ${10}^{-9}$ to $10^{-17}\ \mathrm{eV} \ \mathrm{GeV}^{-2}$ for a single XFEL probe shot.
We found out in the two-dimensional model that phase matching mechanism plays an important role to extend the search bounds.
With proper transverse profile designs of the background or the probe fields, it is possible to concentrate the signal to a certain ALP mass band.
Several kinds of background schemes like target-normal-sheath electrodes field and laser-plasma bubble wakefield are discussed.
The two schemes are also briefly compared. The electrodes scheme is more viable and repeatable, and the bubble scheme has lower disturbance.


Comparing to other photon regeneration experiment plans for $m_a$ below $10^3$ eV, the schemes described in this paper only require one step of conversion and save the expected signals from a second step of feeble interaction.
The approaches to detect the change of the probe's polarization are also highly accurate.
In conclusion, the background field provided by laser-plasma interaction, which is one of the strongest fields people can obtain in a laboratory, has a strong potential application in future ALP searching experiments.

\ack{This work was supported by the Ministry of Science and Technology of the People's Republic of China (Grant No. 2018YFA0404803 and 2016YFA0401102), the Strategic Priority Research Program of the Chinese Academy of Sciences (Grant No. XDB16), and the National Natural Science Foundation of China (Grant No. 11935008).

S Huang would like to express his thanks to Mr Itay Bloch, Drs Sida Lu, Chen Sun, and Ishay Pomerantz from Tel Aviv University, and Drs Oleksandr Borysov and Arka Santra from the Weizmann Institute of Science for inspiring discussions.
}

\clearpage

\section*{References}
\bibliographystyle{iopart-num}
\bibliography{axion_laser}

\clearpage
\appendix
\section{ALP wave function}
\setcounter{section}{1}
\label{sec:appendix1}
The detailed calculation steps solving \eref{eq:kleingordon} are shown here.

The method to obtain \eref{eq:axion1} is called ``coupled-wave equation'' in nonlinear optics.
Under one-dimensional constrain given in \sref{sec:planar}, \eref{eq:kleingordon} turns into
\begin{equation}
	(\partial_t^2 - \partial_x^2 - \nabla_\perp^2 + m^2_a) \psi = g p E_0 B_0 \exp{\{-i(\omega t - k x)\}}.
\end{equation}
A near-field planar axion field is assumed as $\psi_\mathrm{I} = h(x) \exp{\{-i(\omega t - k_a x)\}}$ with a slowly-varying amplitude $h(x)$ that satisfies
\begin{equation}
    2 i k_a h' + h'' = -g p E_0 B_0 \exp{\{i(k-k_a)x\}},
\end{equation}
which is the coupled-wave equation if $|k_a h'| \gg |h''|$. An integration over the interacting region with length $L$ gives
\begin{equation}
\fl
    h(L) = \frac{i g p E_0 B_0}{2 k_a} \int_{-L/2}^{L/2}{\exp{\{i(k-k_a)x\}}}=\frac{i g p E_0 B_0 L}{2 k_a} \mathrm{sinc}\left \{\frac{1}{2}(k-k_a)L \right \}
\end{equation}
and leads to \eref{eq:axion1}. The integral limits are chosen to ensure $h(L)$ a pure-imaginary amplitude. A different pair of the limits can only give an overall phase shift which makes no change to the axion production unless the carrier-envelope phase should be considered.

Now we include the transverse differential term. In cylindrical coordinates with \sref{sec:2b} settings, \eref{eq:kleingordon} turns into
\begin{equation} \label{eq:a5}
	(\partial_t^2 - \partial_x^2 - \partial_r^2 -\frac{1}{r}\partial_r + m^2_a) \psi = g p E_0 B_0 f(r) \exp{\{-i(\omega t - k_x x)\}}
\end{equation}
which is solvable using Bessel--Fourier spectrum methods.
The idea is expanding \eref{eq:a5} inside a cylinder with radius of $\rho$ and Dirichlet boundary condition on the side into a series of independent equations using the complete orthonormal basis of Bessel $\mathrm{J}_0$ functions like
\begin{equation}
		f(r)=\sum_m{f_m \mathrm{J}_0 (k_m r)},\quad 0 \le r < \rho
\end{equation}
\begin{equation}
	f_{m} = \frac{2}{\rho^{2}[\mathrm{J}_{1}(k_{m}\rho)]^{2}}
		\int_{0}^{\rho}{f(r) \mathrm{J}_{0}(k_{m}r) r\mathrm{d}r},
\end{equation}
where $k_m = u_m/\rho$ and $u_m$ is the $m$th zero point of Bessel $\mathrm{J}_0$ function.
There is an upper limit of $m$ to ensure $\omega \ge k_m$.
Assuming $\psi_\mathrm{II}(t,x,r) = \sum_m{\psi_{\mathrm{II}m}}$ and $\psi_{\mathrm{II}m}=h_m(x) \mathrm{J}_0(k_m r) \exp{\{-i(\omega t - k_a x)\}}$, one can find that the slowly-varying amplitude $h_m(x)$ satisfies
\begin{equation}
	k_m^2 h_m - 2 i k_a h'_m = g p E_0 B_0 f_m \exp{\{i(k_x - k_a) x\}}
\end{equation}
which has a pure-imaginary-amplitude integral as
\begin{equation}
    h_m(L) = \frac{i g p E_0 B_0 L f_m}{2 k_a} \mathrm{sinc}\left \{\frac{1}{2}(k_x-k_a-\frac{k_m^2}{2k_a})L \right \}
\end{equation}
that leads to \eref{eq:axion2}. Besides on the transverse cross-section, components $\psi_{\mathrm{II}m}$ are orthonormal to each other because
\begin{equation}
\fl
    \int_0^\rho {\psi_{\mathrm{II}m}\psi^*_{\mathrm{II}n} r\mathrm{d}r}
    \propto f_m f_n \int_0^\rho {\mathrm{J}_0(k_m r)\mathrm{J}_0(k_n r) r\mathrm{d}r}
    = \frac{\mathrm{\delta}_{mn}}{2} f_m f_n \rho^2 [\mathrm{J}_1(k_m \rho)]^2.
\end{equation}
Hence, the total axion production in cylindrical model is the sum of all the contributions from each component.

\section{Possible noise sources}
\label{sec:appendix2}
Unlike the light-by-light scattering in vacuum, the experimental schemes discussed in this paper are based on the background fields driven by laser-plasma interaction.
When the high-power laser hits on materials, including the plasmas as target or the chamber walls, broad band radiations will be created.
It has been widely noticed by the experimentalists that in an intense laser-plasma interaction, high dose of bremsstrahlung radiations in keV waveband are easily observed and have a huge impact on the instrumentation.
During light-matter interaction, the electrons are freed and accelerated by high-power laser to relativistic regime over MeV and then decelerated by the cold part of the environment, and emit photons from keV to MeV.
Unfortunately until so far, experimental reports in in over-10-PW regime are rare to find.
Here we try to scale up the radiation doses obtained at TW to sub-PW regime using the dependency between hot electron temperature and the laser intensity.
According to the studies on 2014 SLAC-MEC experiments using 25 TW laser \cite{Liang2015,Liang2016}, the hot electron temperature $T_\mathrm{h}$  was proportional to the square-root of laser intensity.
In a 40 TW laser experiment at HI Jena, the hot plasma in Ti bulk was created and the x-ray spectra were studied \cite{Rosmej2018}.
The bremsstrahlung spectrum for 20 keV hot electron showed that about $10^{8}$ photons were observed in all $4\pi$ solid angle.
The most recent experiment with 200 TW Vulcan laser showed that $10^{15}$ photons over all spectrum were generated in Si plasma, while 40\% of energy were emitted in silicon's spectral lines between 1.7 to 2.5 keV \cite{Ryazantsev2021}.

For 100 PW laser, the plasma will be over 10 to 30 times hotter.
Nevertheless, the nature of bremsstrahlung photon's profile and the mechanism of x-ray polarizer and analyser as well as the coherency of XFEL photons provide the feasibility to reject the most of photon pollution into the probe, which has been mentioned in \sref{sec:discussions}.
Some extra degrees of freedom can significantly improve the signal-to-noise ratio.
It is also possible to ``encrypt'' the beams with transverse modes, which has been proved practical in the study of quantum information \cite{Wang_2012}.
However, the impact on the probe introduced by intense laser-plasma interaction requires detailed studies to optimize the geometric setup of the final experiment design.
The materials of target and chamber must be selected carefully to exclude their element's spectral lines falling into the probe's frequency, for instance at 1 keV.

The other source of noise could be the ALP interacting with electrons in the presence of low or high-intensity laser (or coherent photon beam), or constant cross field \cite{King_18}.
In our schemes, the electron-ALP interaction can take place in the all three fields: (i)~the background field, (ii) the pump laser field, and (iii) the probe XFEL field.
In the electrode scheme, electrons are pushed out of target perpendicularly to its surface, while the probe is coming parallelly to the target's surface.
ALP created by the ``$ae$'' process has a momentum roughly along the direction of the relativistic electrons, which will not be accepted as a false signal if the ALP decays into photons through ``$a\gamma\gamma$'' process.
But in the bubble scheme, the pump laser is pushing hot electrons ahead of the bubble, while electrons can be injected into the bubble where both the background and XFEL fields are there.
In the second scheme, the electrons can emit an ALP through a Compton-like process, and the ALP may decay into photons that are possible to be identified as a false signal.

However, this event will happen only in a very rare case.
Firstly, the two-step process is proportional to the square of the product of two quite small coupling coefficients $g_{a\gamma\gamma}$ and $g_{ae}$, if the ALP has a long enough lifetime.
Secondly, the ``$ae$'' process will generate ALPs with a continuous energy spectrum, and only a few could decay into photons with the right energy.
Thirdly, pseudoscalar ALP generated in the Compton-like process has a non-zero emitting angle, and will be rejected by the analyser with limited acceptance.
Fourthly, in ALP search, injection of electron in the bubble should always be avoided to keep the background electric field from being reduced.
This is unlike the laser-wakefield accelerators that have a delicately designed target density to make electrons injection into the wakefield bubble as effortless as possible while keep them in the rear part of the bubble.
In this sense, the number of injected electrons that could turn into ALP will be limited.
And lastly, the XFEL beam with $10^{12}$ to $10^{14}$ photons are capable to resist non-polarised noise, as mentioned above and in \sref{sec:discussions}.
Even the suitable ALPs actually turn into polarized x-ray photons under the influence of the background field, it will only be the evidence of ALP's existence.

As a summary, noises from Bremsstrahlung or ALPs generated via a Compton-like process during the laser-plasma interaction need to be studied carefully by experimental experts.
But because the experimental schemes are taking place in a very narrow bandwidth and acceptance angle (eV-mrad), most of the noises can be rejected from being misidentified as signal, given the instrumentation not being destroyed by radiations induced by the unprecedented 100 PW laser.

\end{document}